\documentclass[aps,prb,twocolumn,amssymb,floatfix]{revtex4}  

\usepackage{graphicx}

\renewcommand{\vec}[1]{{\mathbf #1}}

\begin{document}

\title{Dynamic instability of speckle patterns \\ in nonlinear random media}

\author{S.E. Skipetrov}

\affiliation{Laboratoire de Physique et Mod\'elisation des Milieux Condens\'es\\
CNRS, 38042 Grenoble, France}

\begin{abstract}
Linear stability analysis of speckle pattern resulting from
multiple, diffuse scattering of coherent light waves in random media with intensity-dependent refractive index
(noninstantaneous Kerr nonlinearity) is performed.
The speckle pattern is shown to become unstable with respect to dynamic perturbations
within a certain frequency band, provided that nonlinearity exceeds some frequency-dependent threshold. Although
the absolute instability threshold is independent of the response time of nonlinearity, the latter significantly
affects speckle dynamics (in particular, its spectral content) beyond the threshold. Our results suggest that
speckle dynamics becomes chaotic immediately beyond the threshold.


\begin{center}
Accepted for publication in \textit{J. Opt. Soc. Am. B} \copyright~2003 Optical Society of America, Inc.
\end{center}
\end{abstract}

\maketitle 

\section{Introduction}
\label{secintro}

Recent years have been marked by a considerable progress in understanding of
multiple scattering of light in random media in the so-called mesoscopic regime,
when the coherence length $L_{\mathrm{coh}}$ of the incident light exceeds both the size
of the disordered sample $L$ and the absorption length $L_{\mathrm{a}}$, and
$L_{\mathrm{coh}}$, $L$, $L_{\mathrm{a}} \gg \ell$, where $\ell$ is the mean free path
\cite{rossum99,sebbah01,bartsergey}. In the mesoscopic
regime, interference of multiply scattered waves
gives rise to a complicated and seemingly random spatial distribution of light intensity
called `speckle pattern' or just `speckle' for brevity.
Speckles exhibit a number of interesting properties,
such as short- \cite{shapiro86} and long-range \cite{stephen87,zyuzin87,pnini89,albada90,genack90,sebbah02,frank97,shapiro99,skip00c0}
correlations (and related memory effect \cite{freund88} and `universal conductance
fluctuations' \cite{frank98}), and enhanced sensitivity to small displacements of scatterers
\cite{maret87,pine88,berk91}.
As a practical matter, multiply scattered light finds applications 
in optical characterization of turbid materials
(e.g., diffusing-wave spectroscopy \cite{maret87,pine88} and optical microrheology \cite{frank03}),
medical diagnostics \cite{yodh95}, design of random laser sources \cite{wiersma96,wiersma01},
and cryptography \cite{pappu02}.

Most of the available work on multiple light scattering and speckles has concentrated
on {\em linear\/} random media, where the polarization of the medium depends linearly
on the electric field of the light wave. This regime is most often realized in practice
since optical nonlinearities are known to be (generally) weak \cite{bloem96,boyd02}. At the same time,
nonlinear optical systems are known to exhibit much more complex and reach behavior than
their linear counterparts \cite{bloem96,boyd02,gibbs85,arecchi99,voron99},
and it would be therefore important for both fundamental
research and applications to advance our understanding of optical phenomena in
{\em nonlinear\/} random media. Up to now, the interplay of randomness and nonlinearity
has been studied mainly in one-dimensional systems, where
numerical methods proved to be rather efficient (see, e.g., Refs. \onlinecite{sukh01} and
\onlinecite{trom03} for representative bibliography). 
Because waves are known
to behave quite differently in systems of different dimensionalities, 
one-dimensional results are not readily generalizable to the three-dimensional case.
In three-dimensional nonlinear random media, certain attention has been paid to the behavior of {\em average\/}
transmission characteristics in the contexts of
the so-called self-transparency effect \cite{alt86} and optical power limitation \cite{sebbah01b},
whereas available studies of {\em fluctuations\/} of multiply scattered waves
are rather limited \cite{agran91,kravtsov91,boer93,wonderen94,bressoux00,spivak00}.

Recently, a new phenomenon --- dynamic instability of speckle patterns
in random media with intensity-dependent refractive index (Kerr nonlinearity) --- has been theoretically
predicted \cite{skip00,skip01}.
Theoretical study of speckle dynamics beyond the instability threshold has been
initiated in Ref.\ \onlinecite{skip03a} using a newly developed dynamic Langevin approach.
It has been found that speckle pattern is likely to exhibit a transition
from a stationary state to the chaotic dynamics upon increasing the strength of nonlinearity.
The analysis of Ref.\ \onlinecite{skip03a} suffers, however, from numerous limitations, making it
hardly applicable to real experimental situations. Namely, the nonlinearity is assumed to be
instantaneous (nonlinearity response time $\tau_{\mathrm{NL}} = 0$), only the limit of
very large sample size ($L/\ell \gg k \ell$, where $k = 2\pi/\lambda$ and
$\lambda$ is the wavelength of light) is treated, and
the typical time scale $\tau$ of speckle dynamics beyond the instability threshold is assumed to be quite
large: $\tau \gg (\ell/c) (k \ell)^2$, where $c$ is the speed of light in the average medium.
 Meanwhile, noninstantaneous nature of nonlinearity
is known to play an important role in nonlinear dynamics of optical systems
\cite{ikeda80,naka83,silber82,silber84,soljacic00,kip00},
the condition $L/\ell \gg k \ell$ may be difficult to fulfill in practice,
and imposing severe conditions on $\tau$ in the beginning of analysis is undesirable since $\tau$ is
one of the quantities to be found as a result.

In the present paper, we apply the dynamic Langevin approach introduced in Ref.\ \onlinecite{skip03a} to
perform a detailed study of the onset of speckle pattern instability and
speckle dynamics beyond the instability threshold, relaxing the above limitations \cite{skip03b}.
We show that the instability threshold is actually lower than found in Ref.\ \onlinecite{skip03a},
coinciding with the result obtained in Refs.\ \onlinecite{skip00} and \onlinecite{skip01}
in a completely different way. Dynamics of speckle pattern beyond the instability threshold appears
to be very sensitive to relations between $k \ell$ and $L/\ell$, and $\tau_{\mathrm{NL}}$ and
$T_{\mathrm{D}}$, where $T_{\mathrm{D}} = L^2/D$ is the time required for multiply scattered light to
propagate through a random sample of size $L$, and $D$ is the photon diffusion constant.
Although we expect the speckle dynamics to be chaotic beyond the threshold independent of particular
values of above parameters, the spectral content of this dynamics and, in particular, the maximal excited frequency
and Lyapunov exponents, are found to be sensitive to the ratios
 $k \ell/(L/\ell)$ and $\tau_{\mathrm{NL}}/T_{\mathrm{D}}$.
Finally, we comment on the accuracy of our theory by identifying physical processes neglected in the calculation
and discuss the conditions of possible experimental observation of the instability phenomenon for light
in nonlinear random media.

\section{Theoretical model}
\label{model}

We consider multiple scattering of monochromatic (frequency $\omega$) light in
a non-absorbing sample of random medium (typical size $L$) with weak disorder
($k \ell \gg 1$) and intensity-dependent dielectric constant.
For simplicity, and taking into account that multiple scattering depolarizes the light, we apply
scalar approximation from the beginning. The problem then reduces to the following scalar nonlinear
wave equation:
\begin{eqnarray}
&&\left\{ \nabla^2 -
\frac{1}{c^2} \frac{\partial^2}{\partial t^2} \left[
1 + \delta\varepsilon(\vec{r}) +
\Delta\varepsilon_{\mathrm{NL}}(\vec{r}, t) \right]
\right\} E(\vec{r}, t) \nonumber \\
&& \hspace{4cm} = J(\vec{r}, t),
\label{weq}
\end{eqnarray}
where $E(\vec{r}, t)$ is the wave amplitude,
$J(\vec{r}, t) = J_0(\vec{r}) \exp(-i \omega t)$
is a monochromatic source term,
$\delta\varepsilon(\vec{r})$
is the fluctuating part of the linear dielectric constant normalized by its average value
$\left< \varepsilon \right>$,
and the nonlinear part of the dielectric constant
$\Delta\varepsilon_{\mathrm{NL}}(\vec{r}, t)$ (normalized by $\left< \varepsilon \right>$ as well)
depends on the intensity $I(\vec{r}, t) = \left| E(\vec{r}, t) \right|^2$ of the wave.
For $\delta\varepsilon(\vec{r})$ we adopt the Gaussian white-noise model:
$\left< \delta\varepsilon(\vec{r}) \delta\varepsilon(\vec{r}_1) \right> =
4 \pi/(k^4 \ell) \delta(\vec{r} - \vec{r}_1)$. It will be clear from the following that
the specific model used for $\delta\varepsilon(\vec{r})$ should not affect the main results of
the paper significantly because they originate from the long-range diffusion of light in the
random sample and not from the short-range properties of dielectric constant fluctuations.

If the random medium were linear [$\Delta\varepsilon_{\mathrm{NL}}(\vec{r}, t) \equiv 0$],
the time-dependence of $E$ in Eq.\ (\ref{weq}) would be harmonic
[inasmuch as the source term  $J(\vec{r}, t)$ is harmonic]:
$E(\vec{r}, t) = E_0(\vec{r}) \exp(-i \omega t)$,
and hence $I = \left| E_0(\vec{r}) \right|^2$ would be independent of $t$.
Due to nonlinearity, however, the optical system under study can exhibit instabilities resulting in
spontaneous (periodic or chaotic) dynamics of $I$,
despite the fact that the incident radiation [$J_0(\vec{r})$], nonlinear sample [$\delta\varepsilon(\vec{r})$],
and conditions of light propagation remain unchanged (i.e., are time-independent).
Instability phenomena are widely encountered in nonlinear optics
\cite{gibbs85,arecchi99,voron99,ikeda80,naka83,silber82,silber84,soljacic00,kip00}.
Allowing for spontaneous dynamics in our random optical system (but in no way imposing its existence {\em a priori\/}),
we keep the time dependence of $I$ and $\Delta\varepsilon_{\mathrm{NL}}$
in what follows, assuming, however, that this dependence is relatively slow, i.e. that neither $I$ nor
$\Delta\varepsilon_{\mathrm{NL}}$ do not change significantly on the time scale of the
mean free time $\ell/c$.
It will be seen from the following that
spontaneous wave dynamics in nonlinear random media is indeed possible for a sufficiently strong nonlinearity
and that a regime exists when the characteristic time scale of this dynamics is much larger than $\ell/c$.

Due to the randomness of $\delta\varepsilon(\vec{r})$, $E(\vec{r}, t)$ and $I(\vec{r}, t)$ are random functions
of $\vec{r}$ as well (speckle pattern). We will be therefore interested in their statistical moments
rather than in their specific $\vec{r}$- and $t$-dependences for some particular realization
of $\delta\varepsilon(\vec{r})$, an approach common in statistical optics
\cite{rossum99,sebbah01,bartsergey,shapiro86,stephen87,zyuzin87,pnini89,albada90,frank97,freund88,frank98,maret87,pine88,berk91,frank03}.
In particular, we will analyze the correlation function of intensity fluctuations
$\left< \delta I(\vec{r}, t)  \delta I(\vec{r}_1, t_1) \right>$, where  
$\delta I(\vec{r}, t) = I(\vec{r}, t) - \left< I(\vec{r}) \right>$, angular brackets denote averaging
over realizations of disorder $\delta \varepsilon(\vec{r})$, and $\left< I(\vec{r}) \right>$ is
the average intensity that we assume time-independent.
In linear media, spatial variations of $\left< I(\vec{r}) \right>$ are known to be smooth;
$\left< I(\vec{r}) \right>$ changes significantly on the scale of the sample size $L$ but not on the
scale of the mean free path $\ell$. In the following we assume that this smoothness of $\left< I(\vec{r}) \right>$
is preserved in a nonlinear medium. To further simplify the analysis, we assume that
the nonlinearity is weak enough to cause no significant modification of the mean free path $\ell$.
It can be shown \cite{skip03c} that this requires $\left| \Delta \varepsilon_{\mathrm{NL}}(\vec{r}, t) \right|^2 k \ell \ll 1$.
On the one hand, if this condition is violated, our analysis still remains at least qualitatively valid,
provided that the mean free path $\ell$ is recalculated with account for nonlinearity of the medium.
On the other hand, the condition $\left| \Delta \varepsilon_{\mathrm{NL}}(\vec{r}, t) \right|^2 k \ell \ll 1$ does not really
limit the validity of our main results because, as will be seen from the following, instability of speckle
pattern is expected to develop for much weaker nonlinearities than those required to violate the above condition.

\begin{figure}
\includegraphics[width=8cm]{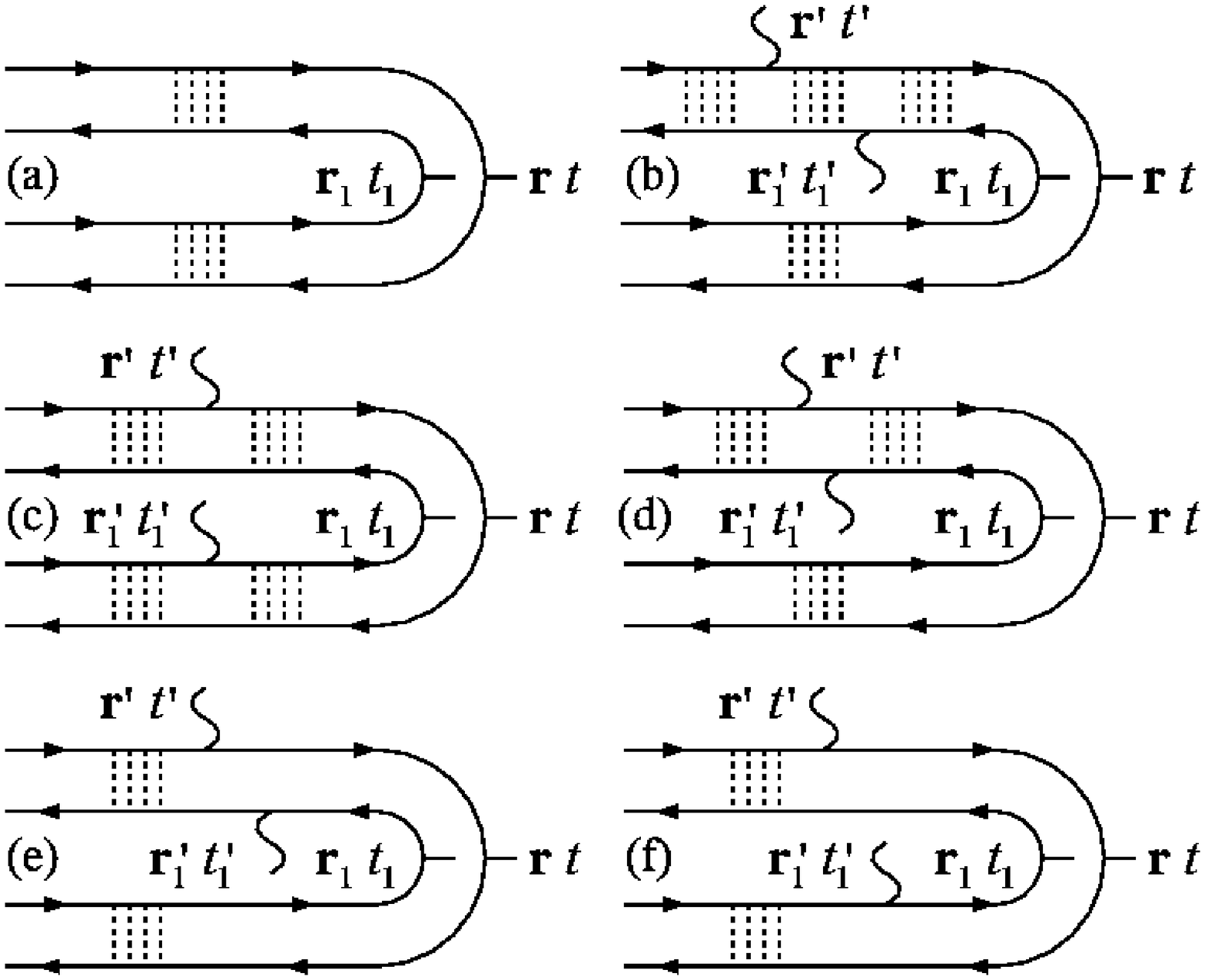}
\caption{
(a) Diagrammatic representation of the correlation function
$\left< j_{\mathrm{ext}}^{(i)} (\vec{r}, t) j_{\mathrm{ext}}^{(j)} (\vec{r}_1, t_1) \right>$
of Langevin currents $\vec{j}_{\mathrm{ext}}(\vec{r}, t)$.
(b)--(d) Diagrams contributing to the correlation function 
$\left< q^{(i)} (\vec{r}, \vec{r}^{\,\prime}, \Delta t)
q^{(j)} (\vec{r}_1, \vec{r}^{\,\prime}_1, \Delta t_1) \right>$
of random response functions $\vec{q}(\vec{r}, \vec{r}^{\,\prime}, \Delta t)$.
(e) and (f) are examples of diagrams neglected in our analysis.
Solid lines with right (left) arrows denote retarded (advanced) Green's functions of the linear disordered wave equation.
Dashed lines denote scattering of the two connected wave fields on the same heterogeneity.
Wavy lines are $k_0^2$ vertices.
The diagrams (b)--(f) are obtained by functional differentiation of the diagram (a) with respect to
the dielectric constant of random medium.
}
\label{fig1}
\end{figure}

It is known that in a linear medium, correlation function of intensity fluctuations
$\left< \delta I(\vec{r}, t)  \delta I(\vec{r}_1, t_1) \right>$ comprises short- and long-range contributions.
The short-range (in space) contribution (often denoted by $C_1$), 
\begin{eqnarray}
\left< \delta I(\vec{r}, t) \delta I(\vec{r}_1, t_1) \right>_{\mathrm{short}} &=&
\left| \left< E(\vec{r}, t) E(\vec{r}, t_1) \right> \right|^2 \nonumber \\
&\times& f(\vec{r}-\vec{r}_1)^2,
\label{short}
\end{eqnarray}
where $f(\vec{r}) = [\sin(k r)/(k r)] \exp[-r/(2 \ell)]$,
is large for $\left| \vec{r} - \vec{r}_1 \right| < \ell$ but
decreases to zero rapidly at larger separations between $\vec{r}$ and $\vec{r}_1$ \cite{shapiro86}.
The-long range contribution is much weaker [$\propto (k \ell)^{-2}$] but persists through the whole
random sample (i.e. even for
$\left| \vec{r} - \vec{r}_1 \right| \sim L$) \cite{stephen87,zyuzin87,pnini89}.
It can be theoretically calculated by using the Langevin approach \cite{zyuzin87,pnini89} that amounts
to writing the Langevin equation for  $\delta I(\vec{r}, t)$,
\begin{eqnarray}
\frac{\partial}{\partial t} \delta I(\vec{r}, t)  - D \nabla^2  \delta I(\vec{r}, t) =
- \vec{\nabla} \cdot {\vec j}_{\mathrm{ext}}(\vec{r}, t),
\label{langevin}
\end{eqnarray}
where ${\vec j}_{\mathrm{ext}}(\vec{r}, t)$ are random  Langevin currents that
have zero mean and correlation functions
$\left< j_{\mathrm{ext}}^{(i)} (\vec{r}, t) j_{\mathrm{ext}}^{(j)} (\vec{r}_1, t_1)
\right> = (c^2/3) \delta_{i j} \left< \delta I(\vec{r}, t) \delta I(\vec{r}_1, t_1) \right>_{\mathrm{short}}$.
Here $i, j = x, y, z$ and $j_{\mathrm{ext}}^{(i)}$ denotes the corresponding Cartesian component
of $\vec{j}_{\mathrm{ext}}$.
Short-range correlation function of intensity fluctuations and correlation function of Langevin currents
can be schematically represented by the diagram (a) of Fig.\ \ref{fig1}.
Exact functional dependence of the long-range correlation function
$\left< \delta I(\vec{r}, t) \delta I(\vec{r}_1, t_1) \right>_{\mathrm{long}}$ on the separation
$\Delta \vec{r} = \vec{r} - \vec{r}_1$ between the points $\vec{r}$ and $\vec{r}_1$, as well as
on the time variables, depends on the geometry of disordered sample and source of waves.
In the infinite random medium with
$\left< I(\vec{r}) \right> \approx I_0$ and for $t = t_1$ one finds \cite{zyuzin87,stephen87}
\begin{eqnarray}
\left< \delta I(\vec{r}, t) \delta I(\vec{r}_1, t) \right>_{\mathrm{long}} \sim
\frac{I_0^2}{k^2 \ell \Delta r}.
\label{long}
\end{eqnarray}
The spatial long-range correlation function of intensity fluctuations (often designated as $C_2$) has been measured
experimentally for microwaves in a quasi-one-dimensional disordered waveguide \cite{genack90,sebbah02}. In that case, the
magnitude of  correlation is of the order of $1/g$ (where $g \gg 1$ is the dimensionless conductance) and not of the order of
$(k \ell)^{-2}$ [as in Eq.\ (\ref{long})], but
it exhibits a similar dependence on $\Delta r$ as in three dimensions: $C_2 \propto 1/\Delta r$.
Spatial correlation function of intensity fluctuations is known to have infinite-range components often designated as
$C_3$ (see, e.g., Ref.\ \onlinecite{rossum99} and references therein)
and $C_0$ (see Refs.\ \onlinecite{shapiro99} and \onlinecite{skip00c0}).
We do not take these latter terms into account in what follows because $C_3 \propto (k \ell)^{-4} \ll C_2$ and
$C_0$ takes considerable values only for the case of spatially-localized (`point-like') source \cite{skip00c0},
whereas we assume an extended source of light.

In linear media, Langevin currents  ${\vec j}_{\mathrm{ext}}(\vec{r}, t)$ are very sensitive to changes of
disorder $\delta\varepsilon(\vec{r})$ \cite{berk91}. Generalizing this property,
one concludes that in a nonlinear medium ${\vec j}_{\mathrm{ext}}(\vec{r}, t)$ should be sensitive to 
changes of $\delta\varepsilon(\vec{r}) + \Delta\varepsilon_{\mathrm{NL}}(\vec{r}, t)$ \cite{spivak00},
and, more precisely, to changes of $\Delta\varepsilon_{\mathrm{NL}}(\vec{r}, t)$, since
$\delta\varepsilon(\vec{r})$ is fixed (and hence does not change) in our case.
Modification $\Delta {\vec j}_{\mathrm{ext}}(\vec{r}, t)$ of Langevin currents due to an infinitesimal change
$\Delta[\Delta\varepsilon_{\mathrm{NL}}(\vec{r}, t)]$ of the nonlinear part of dielectric constant
is \cite{skip03a}
\begin{eqnarray}
\Delta \vec{j}_\mathrm{ext} (\vec{r}, t) &=& \int_V \mathrm{d}^3 \vec{r}^{\, \prime} \int_{-\infty}^{t}
\mathrm{d} t^{\, \prime} \,
\vec{q}( \vec{r}, \vec{r}^{\, \prime}, t - t^{\, \prime} )
\nonumber \\
&\times& \Delta[\Delta\varepsilon_{\mathrm{NL}}(\vec{r}^{\, \prime}, t^{\, \prime})],
\label{dj}
\end{eqnarray}
where the spatial integral is over the volume $V$ of the sample, we neglect
the terms of the second and higher orders in $\Delta\varepsilon_{\mathrm{NL}}(\vec{r}, t)$,
and $\vec{q}( \vec{r}, \vec{r}^{\, \prime}, t - t^{\, \prime} )$ are random, sample-specific
response functions describing the sensitivity of Langevin currents $\vec{j}_\mathrm{ext}$ at $\vec{r}$ at time $t$ to
small changes of dielectric constant $\varepsilon = 1 + \delta\varepsilon + \Delta\varepsilon_{\mathrm{NL}}$
at $\vec{r}^{\, \prime}$ at time $t^{\, \prime}$. (In general case,
$\Delta[\Delta\varepsilon_{\mathrm{NL}}(\vec{r}^{\, \prime}, t^{\, \prime})]$
in Eq.\ (\ref{dj}) should be replaced by
$\Delta\varepsilon(\vec{r}^{\, \prime}, t^{\, \prime})$, but we assume that the linear dielectric constant
$1 + \delta\varepsilon(\vec{r})$ is fixed.)
The physical meaning of $\vec{q}$ can be understood by considering a brief `flash' of $\varepsilon$,
$\Delta\varepsilon(\vec{r}, t) \propto \delta(\vec{r} - \vec{r}_0) \, \delta(t - t_0)$, that occurs at time
$t_0$ and is localized at $\vec{r}_0$. According to Eq.\ (\ref{dj}), for $t > t_0$ such a flash will modify the Langevin current
at some distant point $\vec{r}$ by $\Delta \vec{j}_\mathrm{ext} (\vec{r}, t) \propto
\vec{q}( \vec{r}, \vec{r}_0, t - t_0)$. To account for changes of $\varepsilon$ throughout the whole disordered sample
and at all times before the time $t$, one has to sum their associated $\Delta \vec{j}_\mathrm{ext}$, which results
in integration over space and time and leads to Eq.\ (\ref{dj}).

Obviously, the mean value of $\vec{q}$ is zero, while the correlation function of its Cartesian components
is given by a sum of diagrams (b)--(d) of Fig.\ \ref{fig1}:
\begin{eqnarray}
&&\left< q^{(i)}( \vec{r}, \vec{r}^{\, \prime}, \Delta t) q^{(j)}( \vec{r}_1, \vec{r}_1^{\, \prime}, \Delta t_1) \right>
\nonumber \\
&&=
3 \pi D^2 (c^2/\ell) \delta_{ij} \delta(\vec{r} - \vec{r}_1)
\nonumber \\
&&\times \left\{
\left< I(\vec{r}^{\, \prime}) \right> G(\vec{r}^{\, \prime}, \vec{r}_1^{\, \prime}; \Delta t -\Delta t_1)
G(\vec{r}_1^{\, \prime}, \vec{r}; \Delta t_1) \left< I(\vec{r}) \right> \right.
\nonumber \\
&&+ \left. \left< I(\vec{r}_1^{\, \prime}) \right> G(\vec{r}_1^{\, \prime}, \vec{r}^{\, \prime}; \Delta t_1 - \Delta t)
G(\vec{r}^{\, \prime}, \vec{r}; \Delta t) \left< I(\vec{r}) \right> \right.
\nonumber \\
&&- \left. \left< I(\vec{r}^{\, \prime}) \right> G(\vec{r}^{\, \prime}, \vec{r}; \Delta t)
\left< I(\vec{r}_1^{\, \prime}) \right> G(\vec{r}_1^{\, \prime}, \vec{r}; \Delta t_1) \right.
\nonumber \\
&&+ \left. k^2 \ell/(3 \pi D) \delta(\Delta t - \Delta t_1) f^2(\vec{r}^{\, \prime}-\vec{r}_1^{\, \prime})
\left< I(\vec{r}) \right> \right.
\nonumber \\
&&\times \left. \left[ \left< I(\vec{r}^{\, \prime}) \right> G(\vec{r}^{\, \prime}, \vec{r}; \Delta t) +
 \left< I(\vec{r}_1^{\, \prime}) \right> G(\vec{r}_1^{\, \prime}, \vec{r}; \Delta t) \right]
\right\}.
\label{qcorr}
\end{eqnarray}
Each term in Eq.\ (\ref{qcorr}) and a corresponding diagram of Fig.\ \ref{fig1} can be given a simple
physical interpretation. Consider, for example, the first term,
$\left< I(\vec{r}^{\, \prime}) \right> G(\vec{r}^{\, \prime}, \vec{r}_1^{\, \prime}; \Delta t -\Delta t_1)
G(\vec{r}_1^{\, \prime}, \vec{r}; \Delta t_1) \left< I(\vec{r}) \right>$, given by the diagram (b)
of Fig.\ \ref{fig1}. This term contains two wave fields $E$ and two complex conjugate fields $E^*$
that enter the disordered medium from the left. The first pair of $E$ and $E^*$, shown by two lower
solid lines in Fig.\ \ref{fig1}(b), propagate together to the vicinity of points $\vec{r}$ and
$\vec{r}_1$, experiencing scatterings at exactly the same scatterers (this is denoted by dashed horizontal lines).
After the last scattering event, these two fields do not follow the same path anymore: $E$ goes to $\vec{r}_1$
and $E^*$ goes to $\vec{r}$. The second pair of $E$ and $E^*$, depicted by two upper solid lines in Fig.\ \ref{fig1}(b),
propagate together to some intermediate point $\vec{r}^{\, \prime}$ where $E$ experiences scattering on a fluctuation
$\Delta[\Delta\varepsilon_{\mathrm{NL}}(\vec{r}^{\, \prime}, t^{\, \prime})]$ of the nonlinear part of
dielectric constant. The two wave fields then continue together to some other point $\vec{r}_1^{\, \prime}$,
where $E^*$ is scattered by $\Delta[\Delta\varepsilon_{\mathrm{NL}}(\vec{r}_1^{\, \prime}, t_1^{\, \prime})]$.
Finally, the two wave fields make the remaining way from $\vec{r}_1^{\, \prime}$ to the vicinity of points
$\vec{r}$ and $\vec{r}_1$ together and then separate for $E$ to go to $\vec{r}$
and $E^*$ --- to $\vec{r}_1$. Similar interpretation can also be given to other terms of Eq.\ (\ref{qcorr}) and
to the diagrams (c)--(f) of Fig.\ \ref{fig1}.
We note that the last term in Eq.\ (\ref{qcorr})
[the term containing the factor $f^2(\vec{r}^{\, \prime}-\vec{r}_1^{\, \prime})$], given by the
diagram (d) of Fig.\ \ref{fig1},
has been neglected in Ref.\ \onlinecite{skip03a}.  This is only justified in the limit of extremely
large sample size and for very slow speckle dynamics as will be shown below.
Also, it should be noted that there exist other diagrams that contribute to the
correlation function $
\left< q^{(i)}( \vec{r}, \vec{r}^{\, \prime}, \Delta t)
q^{(j)}( \vec{r}_1, \vec{r}_1^{\, \prime}, \Delta t_1) \right>$ but are neglected in our analysis.
We show examples of such diagrams in panels (e) and (f) of Fig.\ \ref{fig1}. These two particular diagrams
are neglected in Eq.\ (\ref{qcorr}) because their contribution to our principal result,
Eq.\ (\ref{p}), is a factor $(\ell/L)^2 \ll 1$ smaller than the contribution of diagrams
(b)--(d). In addition to neglecting non-leading diagrams in Eq.\ (\ref{qcorr}), we note that
the diagrams (b) and (c) will enter further analysis with the assumption
$\left| \vec{r}^{\, \prime} - \vec{r}_1^{\, \prime} \right| > \ell$. Taking into account the
opposite
possibility ($\left| \vec{r}^{\, \prime} - \vec{r}_1^{\, \prime} \right| \leq \ell$) is possible but
would modify our final Eq.\ (\ref{p}) only by additional terms that are a factor $\ell/L \ll 1$
or a factor $(k \ell)^{-1} \ll 1$
smaller than the terms due to $\left| \vec{r}^{\, \prime} - \vec{r}_1^{\, \prime} \right| > \ell$.

A dynamic equation for $\vec{j}_{\mathrm{ext}} (\vec{r}, t)$ can be obtained from Eq.\ (\ref{dj}) by
assuming that the changes of Langevin currents and nonlinear dielectric constant entering into that equation
take place during some time interval $\delta t$. We then divide both sides of the equation 
by $\delta t$ and take the limit of $\delta t \rightarrow 0$. This yields
\begin{eqnarray}
\frac{\partial}{\partial t} \vec{j}_{\mathrm{ext}} (\vec{r}, t) &=&
\int_V d^3 \vec{r}^{\, \prime} \int_{0}^{\infty} d \Delta t \,
\vec{q}( \vec{r}, \vec{r}^{\, \prime}, \Delta t) \nonumber \\
&\times& \frac{\partial}{\partial t}
\Delta \varepsilon_{\mathrm{NL}}(\vec{r}^{\, \prime}, t - \Delta t).
\label{djnl}
\end{eqnarray}
To describe the temporal evolution of $\Delta\varepsilon_{\mathrm{NL}}(\vec{r}, t)$, we employ
a Debye-type relaxation equation,
\begin{eqnarray}
\tau_{\mathrm{NL}} \frac{\partial}{\partial t} \Delta\varepsilon_{\mathrm{NL}}(\vec{r}, t) =
-\Delta\varepsilon_{\mathrm{NL}}(\vec{r}, t) + 2 n_2 \delta I(\vec{r}, t),
\label{debye}
\end{eqnarray}
which is commonly used in nonlinear optics to model the noninstantaneous medium response \cite{bloem96,boyd02,ikeda80,silber82,silber84}.
Here the constant $n_2$ measures the strength of nonlinearity and
the relaxation time $\tau_{\mathrm{NL}}$ characterizes the speed of nonlinear response of the medium.
Note that we have subtracted from  $\Delta\varepsilon_{\mathrm{NL}}(\vec{r}, t)$ in Eq.\ (\ref{debye})
its average value $2 n_2 \left< I(\vec{r}) \right> \ll 1$ that only leads to a small modification of the
average refractive index of random medium and is of no significance in the context of the present study.
Equations (\ref{langevin}), (\ref{djnl}) and (\ref{debye}) form a self-consistent
system. Equation (\ref{langevin}) governs the spatio-temporal evolution
of the intensity fluctuations $\delta I(\vec{r}, t)$ due to the Langevin
currents ${\vec j}_\mathrm{ext}(\vec{r}, t)$, Eq.\ (\ref{djnl}) describes
the distributed feedback mechanism, leading to variations of
${\vec j}_\mathrm{ext}(\vec{r}, t)$ depending on the changes of $\Delta\varepsilon_{\mathrm{NL}}(\vec{r}, t)$,
and Eq.\ (\ref{debye}) relates  $\Delta\varepsilon_{\mathrm{NL}}(\vec{r}, t)$ to
$\delta I(\vec{r}, t) = I(\vec{r}, t) - \left< I(\vec{r}) \right> $.
Note that Eq.\ (\ref{djnl}) is a linearized equation:
only the terms linear in $\Delta\varepsilon_{\mathrm{NL}}(\vec{r}, t) \ll 1$ are kept.
As we show below,
the linearized nature of Eq.\ (\ref{djnl}) may result in the exponential growth of
its solution with time, and in this sense
Eqs.\ (\ref{langevin}), (\ref{djnl}), and (\ref{debye}) are analogous to the equations of
linear stability analysis commonly used to study the stability of nonlinear systems \cite{gibbs85,arecchi99}.
Hence, although Eqs.\ (\ref{langevin}), (\ref{djnl}), and (\ref{debye}) will allow us to study the stability of the speckle pattern and
the typical time scales of spontaneous intensity fluctuations beyond the
instability threshold, they cannot be used to determine the amplitude of these fluctuations.

\section{Linear stability analysis}
\label{stability}

Consider now an infinitesimal periodic excitation of the stationary speckle pattern:
$\delta I (\vec{r}, t) =  \delta I (\vec{r}, \nu)$ $\exp( i \nu t )$,
where $\nu = \Omega - i \Lambda$.
Such an excitation can be either damped or
amplified, depending on the sign of the Lyapunov exponent $\Lambda$.
The value of $\Lambda$ is determined by two competing processes: on the one hand,
diffusion tends to smear the excitation out, while on the other hand, the
distributed feedback sustains its existence.
The mathematical description of this competition is provided by
Eqs.\ (\ref{langevin}), (\ref{djnl}) and (\ref{debye})
that after the substitution of
$\delta I (\vec{r}, t) =  \delta I (\vec{r}, \nu) \exp( i \nu t )$
[and similarly for $\vec{j}_{\mathrm{ext}} (\vec{r}, t)$ and
$\Delta \varepsilon_{\mathrm{NL}}(\vec{r}, t)$] become
\begin{eqnarray}
&&i \nu \delta I(\vec{r}, \nu) - D \nabla^2 \delta I(\vec{r}, \nu) =
- \vec{\nabla} \cdot {\vec j}_\mathrm{ext}(\vec{r}, \nu),
\label{dinu}
\\
&&i \nu \vec{j}_\mathrm{ext} (\vec{r}, \nu) = i \nu 
\int_V \mathrm{d}^3 \vec{r}^{\, \prime} \int_{0}^{\infty} \mathrm{d} \Delta t \,
\vec{q}( \vec{r}, \vec{r}^{\, \prime}, \Delta t)
\nonumber \\
&&\hspace{2cm} \times \Delta\varepsilon_{\mathrm{NL}} (\vec{r}^{\, \prime}, \nu) \exp(-i \nu \Delta t),
\label{jnu}
\\
&&i \nu \tau_{\mathrm{NL}} \Delta\varepsilon_{\mathrm{NL}} (\vec{r}, \nu) =
- \Delta\varepsilon_{\mathrm{NL}} (\vec{r}, \nu) + 2 n_2 \delta I(\vec{r}, \nu).\hspace{8mm}
\label{enu}
\end{eqnarray}
If $\nu = 0$,  Eq.\ (\ref{dinu}) yields the stationary, time-independent component of the
speckle pattern,
Eq.\ (\ref{jnu}) is trivial, and Eq.\ (\ref{enu}) determines the time-independent
part of the nonlinear dielectric constant (more precisely, its
deviation from the average). This case is not of interest for us.
From here on, we will consider the case of $\nu \ne 0$, because we are interested in {\em dynamics\/} of speckle pattern.
We proceed by expressing  $\Delta\varepsilon_{\mathrm{NL}} (\vec{r}, \nu)$ through
$\delta I(\vec{r}, \nu)$ by using Eq.\ (\ref{enu}) and
the correlation function $\left< \delta I(\vec{r}, \nu)  \delta I^*(\vec{r}_1, \nu) \right>$
through $\left< j_{\mathrm{ext}}^{(i)}(\vec{r}, \nu)  j_{\mathrm{ext}}^{(j)*}(\vec{r}_1, \nu) \right>$
by using Eq.\ (\ref{dinu}).
The latter correlation function can, in its turn, be expressed through 
$\left< \delta I(\vec{r}, \nu)  \delta I^*(\vec{r}_1, \nu) \right>$ by using Eq. (\ref{jnu}).
The two resulting equations relating
$\left< j_{\mathrm{ext}}^{(i)}(\vec{r}, \nu)  j_{\mathrm{ext}}^{(j)*}(\vec{r}_1, \nu) \right>$
and $\left< \delta I(\vec{r}, \nu)  \delta I^*(\vec{r}_1, \nu) \right>$ have to be consistent,
which yields an equation for allowed values of $\Omega$ and $\Lambda$. The calculation can
be performed along the same lines as in Ref.\ \onlinecite{skip03a} and yields
\begin{eqnarray}
p &\simeq& \frac{1 + k\ell/(L/\ell)}{h(\Omega T_D, \Lambda T_D) +
k\ell/(L/\ell) g(\Lambda T_D)} 
\nonumber \\
&\times& H(\Omega \tau_{\mathrm{NL}}, \Lambda \tau_{\mathrm{NL}}),
\label{p}
\end{eqnarray}
where $\Delta n = n_2 {\overline{\left< I(\vec{r}) \right>}}$ is a typical value of the nonlinear
modification of refractive index (with bar denoting spatial average) and
the dimensionless function $h$ is defined in Ref.\ \onlinecite{skip03a}
[Eq.\ (A10) of Ref.\ \onlinecite{skip03a} with $\vec{R} = 0$].
Instead of reproducing here a (rather lengthy) expression for $h(x,y)$, we only note that
$h(0,0) = 1$,
$1-h(x, 0) \propto x^2$ for $x \ll 1$ and  $h(x, 0) \propto x^{-1/2}$ for $x \gg 1$.
Functions $H$ and $g$ in Eq.\ (\ref{p}) are defined as $H(x,y) = x^2 + (1+y)^2$ and
\begin{eqnarray}
g(y) &=&
\frac{2}{(2 - 3 e^{-1/2})} \frac{1}{2 y + 1} \left[ 1 - \left( 1+\frac{1}{2} \sqrt{2 y + 1} \right)
\right.
\nonumber \\
&\times& \left. \exp \left(-\frac{1}{2} \sqrt{2 y + 1}\right) \right].
\label{g}
\end{eqnarray}
Finally, in Eq.\ (\ref{p}) 
we have introduced a {\em bifurcation\/} parameter
\begin{eqnarray}
p = \Delta n^2 \left( \frac{L}{\ell} \right)^2 \left( k \ell + \frac{L}{\ell} \right).
\label{pp}
\end{eqnarray}
As will be seen from the following, this parameter controls the stability of 
speckle pattern with respect to weak perturbations.

Different terms in Eq.\ (\ref{p}) have precise physical meaning. Functions $h$ and $g$ describe the distributed
feedback mechanisms due to long-range ($h$) and short-range ($g$) correlations of intensity fluctuations in random
medium. As follows from Eq.\ (\ref{p}), the relative role of these two feedback mechanisms is controlled by the
parameter $k\ell/(L/\ell)$.  The fact that the nonlinear response of
the medium is noninstantaneous is described by the function $H$. 
$H \equiv 1$ for $\tau_{\mathrm{NL}} = 0$ and $H$ increases monotonically with both $\Omega$ and
$\Lambda$ for $\tau_{\mathrm{NL}} > 0$.

\section{Instantaneous nonlinear response}
\label{fast}

\begin{figure}
\includegraphics[width=9.5cm]{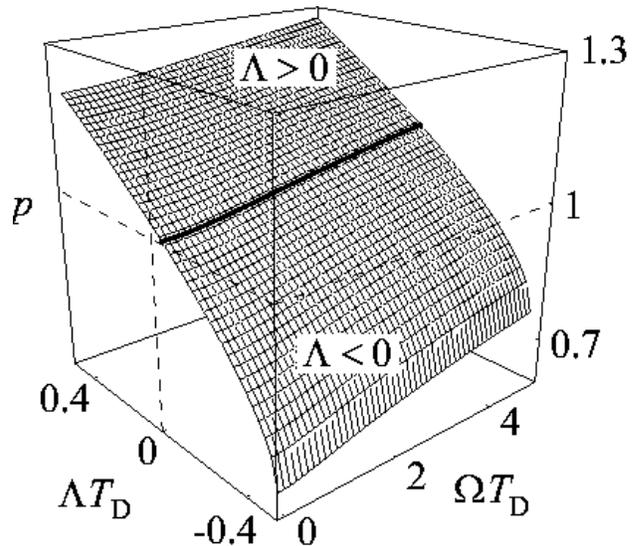}
\caption{
Stability diagram of speckle pattern in a nonlinear disordered medium.
Physically realizable combinations of excitation frequency $\Omega$, Lyapunov exponent $\Lambda$
and bifurcation parameter $p$ belong to the surface shown in the figure.
Thick line corresponds to $\Lambda = 0$ and splits the surface into two parts.
Instability ($\Lambda > 0$) can only develop in the upper part of the surface and hence requires
$p > 1$. In the lower part of the surface, $p < 1$ and $\Lambda < 0$ for all $\Omega$. 
}
\label{fig2}
\end{figure}

We first illustrate the main result of the linear stability analysis performed in the previous section
--- Eq.\ (\ref{p}) --- on the most simple case of instantaneous nonlinearity, assuming
$\tau_{\mathrm{NL}} = 0$.
In this case, 
$H \left( \Omega \tau_{\mathrm{NL}}, \Lambda \tau_{\mathrm{NL}} \right) = 1$
in Eq.\ (\ref{p}). For every value of $p$ and a given $\Omega$ we can then solve Eq.\ (\ref{p})
for $\Lambda$. If $\Lambda < 0$, a weak harmonic (in time) excitation of speckle pattern at the chosen frequency
$\Omega$ is exponentially suppressed and hence disappears after a time $\sim 1/\left| \Lambda \right|$.
If, in contrast, $\Lambda > 0$ our analysis predicts exponential amplification of the excitation and hence
instability of speckle pattern with respect to weak perturbations at frequency $\Omega$.
It follows from Eq.\ (\ref{p}) that $\Lambda < 0$ for all frequencies $\Omega$, if $p < 1$.
We illustrate this in Fig.\ \ref{fig2} where we show a surface plot of
$p$ versus $\Omega$ and $\Lambda$ for a particular case of $k \ell/(L/\ell) = 1$.
A thick line in Fig.\ \ref{fig2} shows a crossection of the surface
by a plane $\Lambda = 0$ and splits the surface into two parts. All points with
$\Lambda > 0$ lie in the upper part of the surface and correspond to $p > 1$. At higher $\Omega$,
larger $p$ is required to get into the upper part of the surface. All points corresponding to $p < 1$ belong
to the lower part of the surface and correspond to $\Lambda < 0$.
When $p$ exceeds $1$, $\Lambda$ becomes positive in a band of frequencies $(0, \Omega_{\mathrm{max}})$
and speckle pattern is hence expected to exhibit instability with respect to low-frequency excitations.
Thus, $p=1$ is the {\em absolute\/} instability threshold. This conclusion agrees with the
result of Refs.\ \onlinecite{skip00} and \onlinecite{skip01} (where the same expression for the absolute
instability threshold has been found by using a different approach) and coincides with the result
of Ref.\ \onlinecite{skip03a} in the limit of $L/\ell \gg k\ell$ considered in that paper.
Equation (\ref{p}) allows us to study dynamics of speckle pattern at $p > 1$ and arbitrary
$L/\ell$, $k\ell \gg 1$.

\begin{figure}
\vspace*{-1.5cm}
\includegraphics[width=10cm]{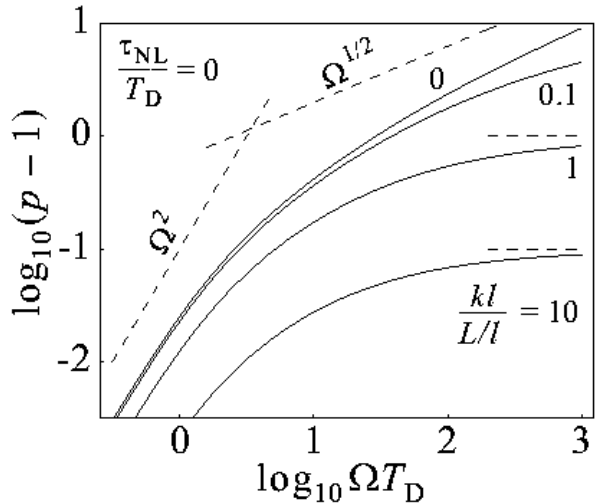}
\caption{
Frequency dependence of instability threshold in a nonlinear disordered medium
with instantaneous Kerr nonlinearity ($\tau_{\mathrm{NL}}/T_{\mathrm{D}} = 0$)
for $k \ell/(L/\ell) \rightarrow 0$ (upper solid curve) and 
$k \ell/(L/\ell) = 0.1$, 1, and 10 (solid curves from top to bottom).
Inclined dashed lines show scaling laws $\Omega^2$ and $\Omega^{1/2}$.
Horizontal dashed lines show asymptotic values of $p$ for large $\Omega$. 
}
\label{fig3}
\end{figure}

Frequency-dependent instability threshold (i.e. the value of $p$ at which $\Lambda = 0$ for a given
$\Omega$) is shown in Fig.\ \ref{fig3} for several values of
$k\ell/(L/\ell)$.
If $k\ell /(L/\ell) \ll 1$ and $\Omega \ll \Omega_1 = T_{\mathrm{D}}^{-1} [(L/\ell)/(k\ell)]^2$,
which corresponds to the limit considered in Ref.\ \onlinecite{skip03a}, 
the denominator of Eq.\ (\ref{p}) is dominated by the first term $h(\Omega T_D, \Lambda T_D)$ and hence
the threshold value of $p$ obeys scaling laws $p-1 \propto (\Omega T_{\mathrm{D}})^2$
for $\Omega \ll T_{\mathrm{D}}^{-1}$ and
$p \propto (\Omega T_{\mathrm{D}})^{1/2}$ for $T_{\mathrm{D}}^{-1} \ll \Omega \ll \Omega_1$ [see
the curves corresponding to $k\ell/(L/\ell) = 0$, $0.1$ and inclined dashed lines in Fig.\ \ref{fig3}].
This regime is governed by the presence of long-range correlations of intensity fluctuations [see
Eq.\ (\ref{long})] and is mathematically described by the diagrams (b) and (c) of Fig.\ \ref{fig1}.
The upper limiting frequency of the band inside which the speckle pattern is unstable
(instability band)
scales as $\Omega_{\mathrm{max}} \propto T_{\mathrm{D}}^{-1} (p-1)^{1/2}$ for $0 < p-1 \ll 1$ and as
$\Omega_{\mathrm{max}} \propto T_{\mathrm{D}}^{-1} p^2$ for $p-1 \gtrsim 1$.
Note that $\Omega_1 = T_{\mathrm{D}}^{-1} [(L/\ell)/(k\ell)]^2 \sim (c/\ell)/(k\ell)^2$ 
exceeds $T_{\mathrm{D}}^{-1}$ for $L/\ell > k\ell$ but is still a factor $(k \ell)^{-2} \ll 1$
smaller than the inverse mean free time $c/\ell$. Therefore, our analysis applies for $\Omega \gtrsim \Omega_1$
as well. In this latter case, the second term, $k\ell/(L/\ell) g(\Lambda T_D)$, in the denominator of Eq.\ (\ref{p})
becomes larger than the first one and we find
$p \simeq 1 + (L/\ell)/(k\ell)$ (see dashed horizontal
lines in Fig.\ \ref{fig3}). As one can see from Fig.\ \ref{fig3}, the curves corresponding
to $k\ell/(L/\ell) = 1$ and $10$ indeed tend to this asymptotic value for $\Omega > \Omega_1$. 
If $p$ exceeds $1 + (L/\ell)/(k\ell)$, $\Lambda > 0$ for {\em all\/} $\Omega$ and speckle pattern becomes unstable
with respect to periodic excitations at {\em all\/} frequencies (in other words, $\Omega_{\mathrm{max}} \rightarrow \infty$).
It is worthwhile to note that this wide-band instability is due to strong short-range correlations between intensity fluctuations in
random media [see Eq.\ (\ref{short})] and mathematically results from the diagram (d) of
Fig.\ \ref{fig1}. The short-range correlations appear to dominate the development of
speckle pattern instability in relatively small random samples ($L/\ell \lesssim k\ell$, but still $L/\ell \gg 1$)
and for high frequencies ($\Omega > \Omega_1$), while in large samples ($L/\ell \gg k\ell$)
and for low frequencies ($\Omega < \Omega_1$) the principal role is played by the long-range correlations.

\begin{figure}
\includegraphics[width=8cm]{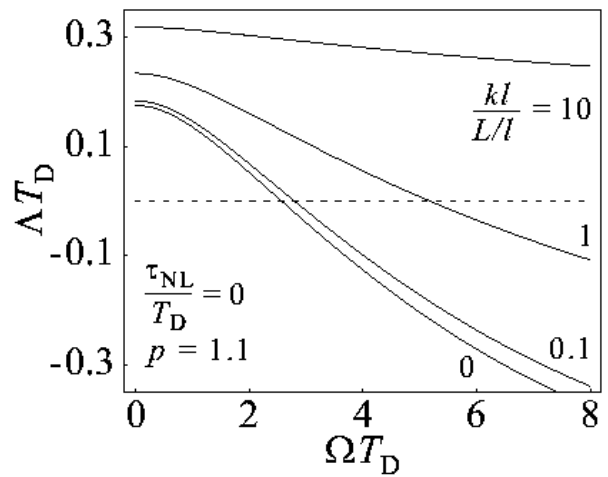}
\caption{
Lyapunov exponent $\Lambda$ versus excitation frequency $\Omega$ (both in units of
$T_{\mathrm{D}}^{-1}=D^2/L$) in a nonlinear disordered medium
with instantaneous Kerr nonlinearity ($\tau_{\mathrm{NL}}/T_{\mathrm{D}} = 0$)
slightly above the absolute instability threshold ($p = 1.1$) 
for $k \ell/(L/\ell) \rightarrow 0$ (lower solid curve) and 
$k \ell/(L/\ell) = 0.1$, 1, and 10 (solid curves from bottom to top).
Horizontal dashed line shows $\Lambda = 0$.
}
\label{fig4}
\end{figure}

Lyapunov exponent $\Lambda$ as a function of frequency $\Omega$ can also be obtained from Eq.\ (\ref{p}).
For a system slightly above the absolute instability threshold ($p = 1.1$), we plot corresponding results
in Fig.\ \ref{fig4} for the same values of $k\ell/(L/\ell)$ as in Fig.\ \ref{fig3}.
In all cases, the maximal value of Lyapunov exponent $\Lambda_{\mathrm{max}}$ is reached for $\Omega \rightarrow 0$
and $\Lambda$ decreases monotonically with $\Omega$.

\section{Noninstantaneous nonlinear response}
\label{slow}

\begin{figure}
\vspace*{-3mm}
\includegraphics[width=8cm]{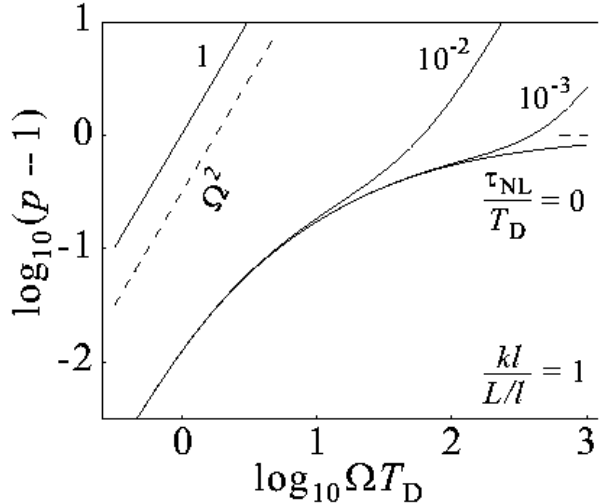}
\caption{
Frequency dependence of instability threshold in a nonlinear disordered medium
with noninstantaneous Kerr nonlinearity ($\tau_{\mathrm{NL}}/T_{\mathrm{D}} = 1$, $10^{-2}$ and  $10^{-3}$,
solid curves from top to bottom) compared with the case of instantaneous nonlinearity
($\tau_{\mathrm{NL}}/T_{\mathrm{D}} = 0$, lower solid curve).
$k \ell/(L/\ell) = 1$ for all curves.
Inclined dashed line shows scaling law $\Omega^2$,
horizontal dashed line shows asymptotic value of $p$ for large $\Omega$ and
$\tau_{\mathrm{NL}}/T_{\mathrm{D}} = 0$. 
}
\label{fig5}
\end{figure}

The case of instantaneous nonlinear response, $\tau_{\mathrm{NL}} = 0$, considered in the
previous section, is, obviously, not very realistic. In this section we extend the analysis of the previous section
to arbitrary $\tau_{\mathrm{NL}}$. For concreteness, we fix the ratio $k\ell/(L/\ell)$ to a realistic value of $1$ and
plot the frequency-dependent instability threshold resulting from Eq.\ (\ref{p}) in Fig.\ \ref{fig5}.
It is readily seen from the figure that the noninstantaneous nature of nonlinearity does not affect the
absolute instability threshold $p = 1$ but modifies significantly its frequency dependence.
We have now several typical times (frequencies) that enter into play: the time $T_{\mathrm{D}}$ needed for
a wave to diffuse through the random sample, the time $\tau_{\mathrm{NL}}$ of nonlinear response, and
the time $\Omega_1^{-1}$, the inverse of the frequency at which the instability threshold saturates at a
constant level in the case of instantaneous nonlinearity (corresponding typical frequencies are
$T_{\mathrm{D}}^{-1}$, $\tau_{\mathrm{NL}}^{-1}$ and $\Omega_1$, respectively).
In the particular case of $k\ell/(L/\ell) = 1$ that we consider, $\Omega_1 = T_{\mathrm{D}}^{-1}$.

Let us first assume that $\tau_{\mathrm{NL}} \ll  T_{\mathrm{D}}$. For $\Omega \ll \tau_{\mathrm{NL}}^{-1}$, the
behavior of instability threshold is then similar to that in the case of instantaneous nonlinearity (see Fig.\ \ref{fig3}
and the lower solid curve of Fig.\ \ref{fig5}).
First, the threshold value of $p-1$ scales as $(\Omega T_{\mathrm{D}})^2$
for $\Omega < T_{\mathrm{D}}^{-1}$, then the curve starts to saturate at a constant level for $\Omega > \Omega_1$
[if $L/\ell > k\ell$ and $T_{\mathrm{D}}^{-1} < \Omega_1$, there is a region $T_{\mathrm{D}}^{-1} < \Omega < \Omega_1$,
where $p \propto (\Omega T_{\mathrm{D}})^{1/2}$].
However, in contrast to the case of instantaneous nonlinearity, this saturation is not complete and is followed by a parabolic
growth of $p$ with $\Omega$ for $\Omega > \tau_{\mathrm{NL}}^{-1}$. Physically, this is due to the fact that
the nonlinear part of dielectric constant $\Delta \varepsilon_{\mathrm{NL}}$ cannot follow such fast dynamics
and the amplitude of its oscillations is strongly suppressed for  $\Omega > \tau_{\mathrm{NL}}^{-1}$. Mathematically,
the growth of the threshold value of $p$ with $\Omega$ for  $\Omega > \tau_{\mathrm{NL}}^{-1}$ originates from the
term $H(\Omega\tau_{\mathrm{NL}}, \Lambda\tau_{\mathrm{NL}})$ in Eq.\ (\ref{p}).
This behavior is illustrated by the solid curves corresponding to $\tau_{\mathrm{NL}} / T_{\mathrm{D}} = 10^{-3}$
and $10^{-2}$ in Fig.\ \ref{fig5}.

If the response time of nonlinearity $\tau_{\mathrm{NL}}$ is of the order or shorter than
$T_{\mathrm{D}}$ (see the upper solid curve in Fig.\ \ref{fig5}), the frequency dependence of instability
threshold becomes dominated by the noninstantaneous nature of nonlinearity at all frequencies. We find
$p-1 \propto (\Omega \tau_{\mathrm{NL}})^2$ as illustrated in Fig.\ \ref{fig5} for
 $\tau_{\mathrm{NL}} / T_{\mathrm{D}} = 1$.

\begin{figure}
\includegraphics[width=8cm]{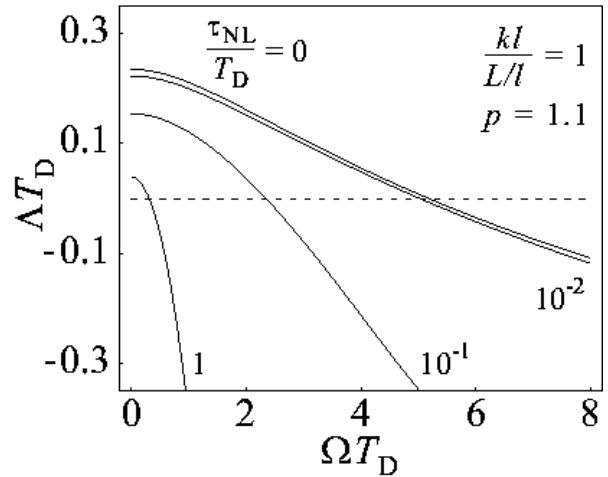}
\caption{
Lyapunov exponent $\Lambda$ versus excitation frequency $\Omega$ (both in units of
$T_{\mathrm{D}}^{-1}=D^2/L$) in a nonlinear disordered medium
with noninstantaneous Kerr nonlinearity
($\tau_{\mathrm{NL}}/T_{\mathrm{D}} = 1$, $10^{-1}$ and  $10^{-2}$,
solid curves from bottom to top), slightly above the absolute instability threshold ($p = 1.1$),
compared to the case of instantaneous nonlinearity
($\tau_{\mathrm{NL}}/T_{\mathrm{D}} = 0$, upper solid curve).
$k \ell/(L/\ell) = 1$ for all curves.
Horizontal dashed line shows $\Lambda = 0$.
}
\label{fig6}
\end{figure}

The frequency dependence of Lyapunov exponent $\Lambda$ found by solving Eq.\ (\ref{p}) slightly
above the absolute instability threshold ($p = 1.1$) and for
$k\ell/(L/\ell) = 1$ is shown in Fig.\ \ref{fig6} for several values of  $\tau_{\mathrm{NL}} / T_{\mathrm{D}}$.
Qualitatively, behavior of $\Lambda$ is similar to that found in the case of instantaneous nonlinearity
(see Fig.\ \ref{fig4} and the upper solid curve of Fig.\ \ref{fig6}), but $\Lambda$ decreases with $\Omega$ faster
than for
$\tau_{\mathrm{NL}} / T_{\mathrm{D}} = 0$. Together with a faster growth of instability threshold
(Fig.\ \ref{fig5}), this narrows the band of frequencies $(0, \Omega_{\mathrm{max}})$ where excitations are subject 
to instability. Instead of becoming infinite for large $p$ as in the case of instantaneous nonlinearity
($\tau_{\mathrm{NL}} = 0$), $\Omega_{\mathrm{max}}$ scales as
$\tau_{\mathrm{NL}}^{-1} p^{1/2}$ for $\tau_{\mathrm{NL}} / T_{\mathrm{D}} \gg 1$.

\section{Path-integral picture of speckle pattern instability}
\label{path}

It is worth noting that most of the peculiarities of the above analysis can be qualitatively understood in the path-integral
picture of wave propagation, especially in the case of instantaneous nonlinearity that we now consider in more detail.

Consider a partial wave traveling along a typical
diffuse path of length $s \sim L^2/\ell$ in a nonlinear disordered medium.
Assume that the path length $s$ can vary slowly with time due to, e.g., an infinitely slow Brownian
motion of scatterers.
A diffuse wave following the same path a time $\Delta t$ later will then acquire a different phase.
The phase difference between the two waves, $\Delta \phi$, can be splited in a sum of
two contributions: a `linear' contribution $\Delta \phi_{\mathrm{L}}$, resulting directly from the
change of path length $s$ and independent of the nonlinear properties of the medium,
and a `nonlinear' contribution $\Delta \phi_{\mathrm{NL}}$ due to the nonlinearity of the medium:
\begin{eqnarray}
\Delta \phi_{\mathrm{NL}} = k n_2 \int_0^{s} ds_1
\Delta I(\vec{r}),
\label{phasenl}
\end{eqnarray}
where $\Delta I(\vec{r})$ denotes the change of intensity during the time $\Delta t$,
we take into account the fact that the nonlinear part of the refractive index is simply
$n_2 I(\vec{r})$ in the considered case of instantaneous nonlinearity, and the integral is along the wave path.
In the limit of weak nonlinearity, it can be shown \cite{skip01,skip03c}
that the mean values of both $\Delta \phi_{\mathrm{L}}$ and $\Delta \phi_{\mathrm{NL}}$ are zeroes, while their
variances are related by
\begin{eqnarray}
\left< \Delta \phi_{\mathrm{NL}}^2 \right> &=&
\left< \Delta \phi_{\mathrm{L}}^2 \right>
k^2 n_2^2 
\nonumber \\
&\times& \int_0^{s} ds_1 \int_0^{s} ds_2
\left< \delta I(\vec{r}_1) \delta I(\vec{r}_2) \right>,
\label{varnl2}
\end{eqnarray}
where both integrals are along the same diffusion path.
Because we have assumed nonlinearity to be weak, we can proceed by perturbation and replace
the correlation function of intensity fluctuations
$\left< \delta I(\vec{r}_1) \delta I(\vec{r}_2) \right>$ in Eq.\ (\ref{varnl2})
by its value in the linear medium, i.e., by a sum of
two contributions: the short-range one [Eq.\ (\ref{short})] and the long-range one [Eq.\ (\ref{long})]. 
To perform integration in Eq.\ (\ref{varnl2}), we replace $\Delta r = \left| \vec{r}_1 - \vec{r}_2 \right|$
in the expression (\ref{short}) for the short-range correlation function by $\Delta s = \left| s_1 - s_2 \right|$,
assuming the
wave path to be ballistic at distances shorter than $\ell$.
In contrast, for $\Delta r > \ell$ the wave path is diffusive, and hence $\Delta r$ in the expression
(\ref{long}) for the long-range correlation function should be substituted by 
$(\Delta s \ell)^{1/2}$.  
Performing then integrations in Eq.\ (\ref{varnl2}) yields
\begin{eqnarray}
\left< \Delta \phi_{\mathrm{NL}}^2 \right> \simeq
p \left< \Delta \phi_{\mathrm{L}}^2 \right>,
\label{varnl3}
\end{eqnarray}
with $p$ the bifurcation parameter defined by Eq.\ (\ref{pp}) with
the first term, $\Delta n^2 (L/\ell)^2 k\ell$, resulting from the substitution of the 
short-range correlation function (\ref{short}) in Eq.\ (\ref{varnl2}) and the second term,
$\Delta n^2 (L/\ell)^3$, --- from the substitution of the long-range correlation function
(\ref{long}).
It follows from Eq.\ (\ref{varnl3}) that at $p > 1$, the phase difference
$\Delta \phi = \Delta \phi_{\mathrm{L}} + \Delta \phi_{\mathrm{NL}}$ is dominated by $\Delta \phi_{\mathrm{NL}}$
and hence one may expect new phenomena (such as instability) to appear. Because we have already shown that
$p = 1$ is indeed the instability threshold, we adopt $\left< \Delta \phi_{\mathrm{NL}}^2 \right> >
\left< \Delta \phi_{\mathrm{L}}^2 \right>$ as the instability condition within our path-integral picture.

The path-integral picture of wave propagation can be also applied to describe the dependence of
instability threshold on the frequency of excitation.  Indeed, assume that speckle
pattern oscillates at some frequency $\Omega$ that (we remind) is assumed to be much smaller than the inverse
mean free time $c/\ell$. At low frequencies $\Omega \ll T_{\mathrm{D}}^{-1}$
we recover Eq.\ (\ref{varnl3}) because the oscillations are very slow and everything happens as if the speed
of light diffusion through the random medium were infinite.
If, in contrast, $\Omega > T_{\mathrm{D}}^{-1}$, situation changes.
In Eq.\ (\ref{varnl2}), the result of integration of the short-range correlation function
(\ref{short}) remains the same as before since this correlation is only non-zero for
$\Delta r < \ell$ and $\Omega^{-1}$ is still much larger than the time needed for light to cover this distance.
In contrast, integration of the long-range correlation function (\ref{long}) now yields a different result.
Indeed, in this case correlation between intensities at
$\vec{r}_1$ and $\vec{r}_2$ can only be non-zero for $\Delta r = \left| \vec{r}_1 - \vec{r}_2 \right|
\leq (D \Omega^{-1})^{1/2} < L$ due to the finite speed of light diffusion in random medium.
Integration in Eq.\ (\ref{varnl2}) is therefore limited to pairs of points $\vec{r}_1$, $\vec{r}_2$
separated not more than by a distance $(D \Omega^{-1})^{1/2}$. This gives
\begin{eqnarray}
\left< \Delta \phi_{\mathrm{NL}}^2 \right> \simeq
\Delta n^2 \left( \frac{L}{\ell} \right)^2
\left[ k \ell + \frac{L}{\ell} \left(\Omega T_{\mathrm{D}} \right)^{-1/2}
\right]
\left< \Delta \phi_{\mathrm{L}}^2 \right>. \hspace{3mm}
\label{varnl4}
\end{eqnarray}
The condition of instability of speckle pattern with respect to excitations at frequency $\Omega$,
$\left< \Delta \phi_{\mathrm{NL}}^2 \right> > \left< \Delta \phi_{\mathrm{L}}^2 \right>$,
can now be written as
\begin{eqnarray}
p \gtrsim \frac{1 + k\ell/(L/\ell)}{(\Omega T_D)^{-1/2} +
k\ell/(L/\ell)}.
\label{p2}
\end{eqnarray}
This coincides with Eq.\ (\ref{p}) for $\Omega > T_{\mathrm{D}}^{-1}$.

The path-integral picture of speckle pattern instability can help to understand how to apply the theoretical model developed
in this paper to realistic experimental situations, when the disordered sample is characterized
by three different size parameters $L_x$, $L_y$, and $L_z$ (in the so-called `slab' geometry, for example, $L_x = L_y \gg L_z \gg \ell$).
In this case, the only size parameter $L$ of our theoretical model should be understood as a size of the part of disordered sample
explored by a typical diffuse path (and hence $L \approx L_z$ in the slab geometry).

\section{Discussion and conclusions}
\label{concl}

Theoretical analysis performed in the present paper indicates that the multiple-scattering speckle pattern
in a nonlinear random medium with intensity-dependent dielectric constant becomes unstable with respect to
weak dynamic perturbations if the nonlinearity exceeds the threshold $p \simeq 1$. This absolute instability threshold
appears to be independent of the nonlinearity response time $\tau_{\mathrm{NL}}$, whereas the frequency dependence
of the threshold (and, in particular, the spectrum of frequencies subject to instability)
is very sensitive to $\tau_{\mathrm{NL}}$.
It is pertinent to note the extensive nature of the bifurcation
parameter $p$. According to Eq.\ (\ref{pp}),
weak nonlinearity can be efficiently compensated by a sufficiently
large sample size $L$ and $p \sim 1$ can be reached even at vanishing
$\Delta n$, provided that $L$ is large enough.
Another remarkable feature of Eq.\ (\ref{pp}) is the independence of $p$ of
the sign of $\Delta n$. The phenomenon of speckle pattern instability is
therefore expected to develop in a similar way for both positive and
negative nonlinear coefficients $n_2$. This is not common for instabilities
in nonlinear optical systems without disorder
\cite{bloem96,boyd02,gibbs85,arecchi99,voron99,ikeda80,naka83,silber82,silber84,soljacic00,kip00}
since the latter instabilities are often related to the self-focusing effect, arising at $n_2 > 0$ only. The instability
discussed in the present paper is of different type and has nothing to do with
self-focusing (which is negligible for $\Delta n^2 k\ell \ll 1$ assumed throughout the paper).
This does not mean, however, that phenomena similar to those discussed
in the present paper
do not exist in homogeneous nonlinear media. In fact, it is easy to see that
Eq.\ (\ref{phasenl}) describes nothing but the self-phase modulation \cite{boyd02}
of the multiple-scattering speckle pattern.
The development of self-phase
modulation in a random medium, however, appears to be rather different from that in the
homogeneous case. Indeed, in a homogeneous medium of size $L$ the nonlinear phase shift
is deterministic and $\phi_{\mathrm{NL}} = k n_2 I_0 L$ for a wave
of intensity $I_0$, whereas in the case of diffuse multiple scattering
$\phi_{\mathrm{NL}}$ is random with the average value
$\left< \phi_{\mathrm{NL}} \right> \simeq k n_2 I_0 (L^2/\ell)$ and variance
$\left< \phi_{\mathrm{NL}}^2 \right> -
\left< \phi_{\mathrm{NL}} \right>^2 \simeq p$.
The threshold of the speckle pattern instability $p \simeq 1$ can be therefore viewed as simply
the point where the sample-to-sample fluctuations of the nonlinear phase
shift become of order unity.

It is worth while to comment on the connection of the problem considered in the present paper to
the recent results on modulational instability (MI) of incoherent light beams in homogeneous nonlinear media \cite{soljacic00,kip00}.
In this case, a 2D speckle pattern (speckled beam) with a controlled degree of spatial and temporal coherence is
`prepared' by sending a coherent laser beam on a rotating diffuser. The beam is then incident
on a homogeneous nonlinear medium (inorganic photorefractive crystal).
Pattern formation and `optical turbulence' are observed if the intensity of the beam exceeds a specific
threshold, determined by the degree of coherence of the beam.
Although the study presented in this paper is also concerned with
the instability of speckle patterns in nonlinear media, it
differs essentially from that of Refs.\ \onlinecite{soljacic00} and \onlinecite{kip00}
 at least in the following important aspects:
(a) light loses its spatial coherence
\textit{inside} the medium, due to the multiple scattering on heterogeneities of refractive index,
(b) the temporal coherence of
incident light is perfect (in the case of incoherent MI \cite{soljacic00,kip00},
coherence time of the incident beam is much shorter than the nonlinearity response time
$\tau_{\mathrm{NL}}$),
and (c) the incident light beam is completely destroyed by scattering after a distance 
$\sim \ell$  and light propagation is diffusive in the bulk of the sample.

Let us now say a few words about practical consequences of our results and conditions of possible experimental observation
of the instability phenomenon for light in random media. In a real experiment, infinitely weak dynamic
perturbations of stationary speckle pattern are always present (due to, e.g., various types of noise).
Our analysis predicts that the spectral components of these perturbations falling within the instability band
$(0, \Omega_{\mathrm{max}})$ will be amplified due to the nonlinearity of the medium. Because our analysis is based on
a linearized equation (\ref{djnl}), we cannot predict further development of speckle dynamics with certainty,
but our results
suggest the following scenario.
In general, there are two possibilities for the spontaneous dynamics of speckle pattern beyond the threshold. First, it can be
(quasi-) periodic, i.e. dominated by oscillations at some discrete frequency (or frequencies) $\Omega_i$. Second, it can be
chaotic with a continuous spectrum of intensity fluctuations between some $\Omega_{\mathrm{min}}$ and 
$\Omega_{\mathrm{max}}$. In nonlinear systems where dynamics is periodic, the frequency of oscillations
$\Omega_i$ is often favored already on the level of the linear stability analysis: it is often the
frequency for which the maximum Lyapunov exponent $\Lambda$ is reached. In our case, $\Lambda$ exhibits
monotonic decrease with $\Omega$ (see Figs.\ \ref{fig4} and \ref{fig6}), favoring no
particular frequency. We have therefore no reason to expect the first scenario and the speckle pattern $I(\vec{r}, t)$ is
likely to exhibit chaotic dynamics immediately beyond the instability threshold. The spectrum of chaotic
fluctuations of $I(\vec{r}, t)$ will be continuous and comprised between  $\Omega_{\mathrm{min}} = 0$ and
$\Omega_{\mathrm{max}}$ that we have discussed above.
We note, however, that our analysis assumes diffusion regime of wave propagation
($k \ell \gg 1$ and $L/\ell \gg 1$) and is likely to fail if the scattering is low-order ($L \lesssim \ell$)
or if the regime of Anderson localization is approached ($k \ell \sim 1$). In the latter cases, the first scenario
of instability development can be a plausible option.

Finally, experimental observation of instability of multiple-scattering speckle pattern will require,
first of all, large enough sample size $L$ and as low absorption
as possible (if the macroscopic absorption length
$L_{\mathrm{a}} <  L$, the effective size of the sample will be limited to $L_{\mathrm{a}}$ and one will not
fully exploit the extensivity of instability threshold). In the absence of absorption (or for 
$L_{\mathrm{a}} \gtrsim L$),  $\Delta n_{\mathrm{NL}} \sim 10^{-2}$ (realistic in nematic liquid
crystals) and $L/\ell \sim 20$ will suffice to
get $p \simeq 1$ and reach the instability threshold.
We note that in liquid crystals, optical nonlinearity is due to reorientation of molecules under the
influence of electric field of electromagnetic wave and hence is essentially
noninstantaneous. This emphasizes the importance of including the noninstantaneous
nature of nonlinear response in our analysis.

\end{document}